\documentstyle[11pt,newpasp,twoside,epsf]{article}
\def\edcomment#1{\iffalse\marginpar{\raggedright\sl#1\/}\else\relax\fi}
\reversemarginpar

\def \Lq {L_{\rm {q}}}
\def \Lout {L_{\rm {out}}}

\def \Tc {T_{\rm {c}}}

\def \trec {t_{\rm {rec}}}
\def \tout {t_{\rm {out}}}
\def \Msun {${\rm M_{\odot}}~$}
\def \mdot {\dot {M}}
\def \tpycno {t_{\rm {py}}}
\def \mdotav {\langle \dot {M}\rangle }

\begin{document}
\title {Neutron star cooling in transiently accreting low mass binaries:
a new tool for probing nuclear matter}

\author{Andrea Possenti}
\affil{Osservatorio di Bologna, via Ranzani 1, 40127 Bologna, Italy}
\author{Monica Colpi}
\affil{Univ. di Milano Bicocca, P.zza della Scienza 3, 20126 Milano, Italy}
\author{Dany Page}
\affil{Instituto de Astronom\'{\i}a, UNAM, 04510 M\'{e}xico D.F., M\'{e}xico}
\author{Ulrich Geppert}
\affil{AIP, An der Sternwarte 16, D-14482, Potsdam, Germany}

\begin{abstract}
We explore, using an exact cooling code, the thermal evolution of a neutron 
star undergoing episodes of intense accretion, alternated by long periods 
of quiescence (e.g. Soft X-Ray Transients; SXRTs hereon).
We find that the soft component of the quiescent luminosity of Aql X-1, 
4U 1608-522 and of SAX J1808.4-3658 can be understood as thermal emission 
from a cooling neutron star with negligible neutrino emission. 
In the case of Cen X-4 strong neutrino emission from the inner core 
is necessary to explain the observation: this may indicate that the neutron 
star of Cen X-4 is heavier than 1.4 \Msun. This study opens the possibility 
of using the quiescent emission of SXRTs as a tool for probing the core 
superfluidity in relation to the mass of the neutron star.

\end{abstract}

\section{Introduction}

The neutron stars (NSs) of the SXRTs undergo recurrent surges of X-ray 
activity (due to intense accretion onto the stellar surface)
separated by longer periods of relative {\it quiescence}. The quiescent 
emission, around $10^{32-33}$ erg s$^{-1},$ is characterized by a thermal 
spectrum plus a power law tail at higher energies. Though accretion can not be
excluded as energy source (van Paradijs et al. 1987), observations hint 
in favor of interpreting the thermal component as due to the cooling of the 
old NS, heated during the episodes of intense accretion (Brown, Bildsten 
\& Rutledge 1998, BBR98 hereafter; Rutledge et al. 2000; Campana et al. 1998).
Freshly accreted matter undergoes nuclear burning (Brown 2000)
whose energy is almost instantly lost through the photosphere.
As accretion proceeds recurrently, the NS envelope becomes  
progressively enriched of Fe elements. The old crust is eventually assimilated
into the core and the NS, after a few million years, is 
endowed by a new crust of non-equilibrium matter which
is approaching its lowest energy state through a sequence of pycnonuclear 
reactions (Haensel \& Zdunik 1990a, 1990b). The heat, deposited deeply into 
the crust, emerges in quiescence with a luminosity $\Lq$ (visible between 
outbursts), depending on the balance between photon and neutrino cooling 
with pycnonuclear heating. We here calculate $\Lq$ with an exact cooling 
code to derive the upper theoretical bound on the efficiency of rediffusion 
of the pycnonuclear energy as function of the recurrence time of a SXRT.
We will also prove that charting the temperature of the old hot NS 
in a SXRT is a valuable tool to investigate the properties of nuclear 
matter in its core, in alternative to the study of isolated young cooling NSs. 
As some of these compact objects may have accreted a substantial amount of
mass from the companion (Burderi et al. 1999), this study  
opens also the possibility to explore the interior 
of NSs which can be more massive than the isolated ones.
As an illustration of this approach we compare the results of our calculation
with the observation of the quiescent emission seen in some SXRTs 
(Aql X-1, Cen X-4,  4U 1608-522, EXO 0748-676, Rapid Burster) and in the 
transient X-ray 2.5 ms pulsar SAX J1808.4-3658. The agreement between theory 
and observation is quite remarkable.

\section{Modeling cooling and transient accretion}
\label{sect:model}

We use an exact cooling code which solves the equations of heat 
transport 
\footnote{
The thermal conductivity is as in Colpi et al. (2001). 
We use the calculations of Potekhin, Chabrier and Yakovlev (1997) for a 
cooling star with an accreted envelope to compute 
$\Lq$ and the 
corresponding effective temperature $T_{\rm eff}.$}
and energy conservation in a wholly general relativistic scheme (Page 1998).
The {\sl cooling sources} are neutrino emission (in
the crust and the core) and surface photon radiation.
The {\sl heating source} is accretion-induced production of nuclear energy: 
in a fully replaced crust, 
the bulk of the energy is released by the pycnonuclear fusions
($Q_{\rm{py}}\sim 0.9$ MeV/baryon) in a time-scale 
$\tpycno \sim$ months 
\footnote{We use a heat release function 
\begin{equation} 
{\cal{R}}_{\rm {py}}(t)=
\frac{1}{\tpycno} \;\int_0^t\,\,dt'\mdot_{\rm {p}}(t'){Q_{\rm{py}}
\over m_u}\exp[(t'-t)/\tpycno] 
\end{equation}
representing the energy deposited per unit time at the current time. 
A minor energetic contribution comes from electron captures and neutron 
emissions (Haensel \& Zdunik 1990a) occurring at a rate proportional 
to the instantaneous accretion rate $\mdot_{\rm {p}}(t)$.$m_u$ is the atomic
mass unit.}.

The chemical composition and the equation of state (EOS) of the crust 
are the ones of an accreted crust (Haensel \& Zdunik 1990b) while 
for the core we follow Prakash, Ainsworth \& Lattimer (1988). We include 
the strong suppressing effects of superfluidity (neutron) and 
superconductivity (protons) on both the specific heat and neutrino emission
\footnote{Generically, stars with mass $1.4 \to 1.6$ \Msun have only the 
modified Urca process allowed in their core and suppression by neutron 
pairing is strong,
while  stars with $1.7 \to 1.8$ \Msun have the direct Urca process 
operating with suppression at lower temperatures so that fast neutrino 
cooling does affect significantly their thermal evolution.}.

We model the transient accretion rate $\dot{M}(t)$ (measured at $\infty$)  
with a fast exponential rise, on a time scale $t_{\rm rise}\sim 10$ days, 
reaching a maximum $\dot{M}_{\rm max}$, followed by a power law decay of 
index $\alpha=3.$ Accretion is never turned off but becomes negligible 
at $t_{\rm {out}}=3t_{\rm {rise}}\sim 30$ days (a typical duration 
$t_{\rm {out}}$ of the outbursts in SXRTs). A new exponential rise recurs 
every $t_{\rm rec}$. Thus, the total accreted mass 
during one cycle is $\Delta M=1.08 \,t_{\rm rise}\,\dot{M}_{\rm max}$; the
time averaged accretion rate is $\langle\dot M\rangle=\Delta M/t_{\rm rec};$
and the outburst luminosity (at $\infty$) $L_{\rm out}$ is such that the 
fluence $L_{\rm out} \; t_{\rm out} =\eta \; \Delta M \,c^2,$ with $\eta$ the efficiency. 

\section{Equilibrium temperature and quiescent luminosity}

\begin{figure}
\plottwo{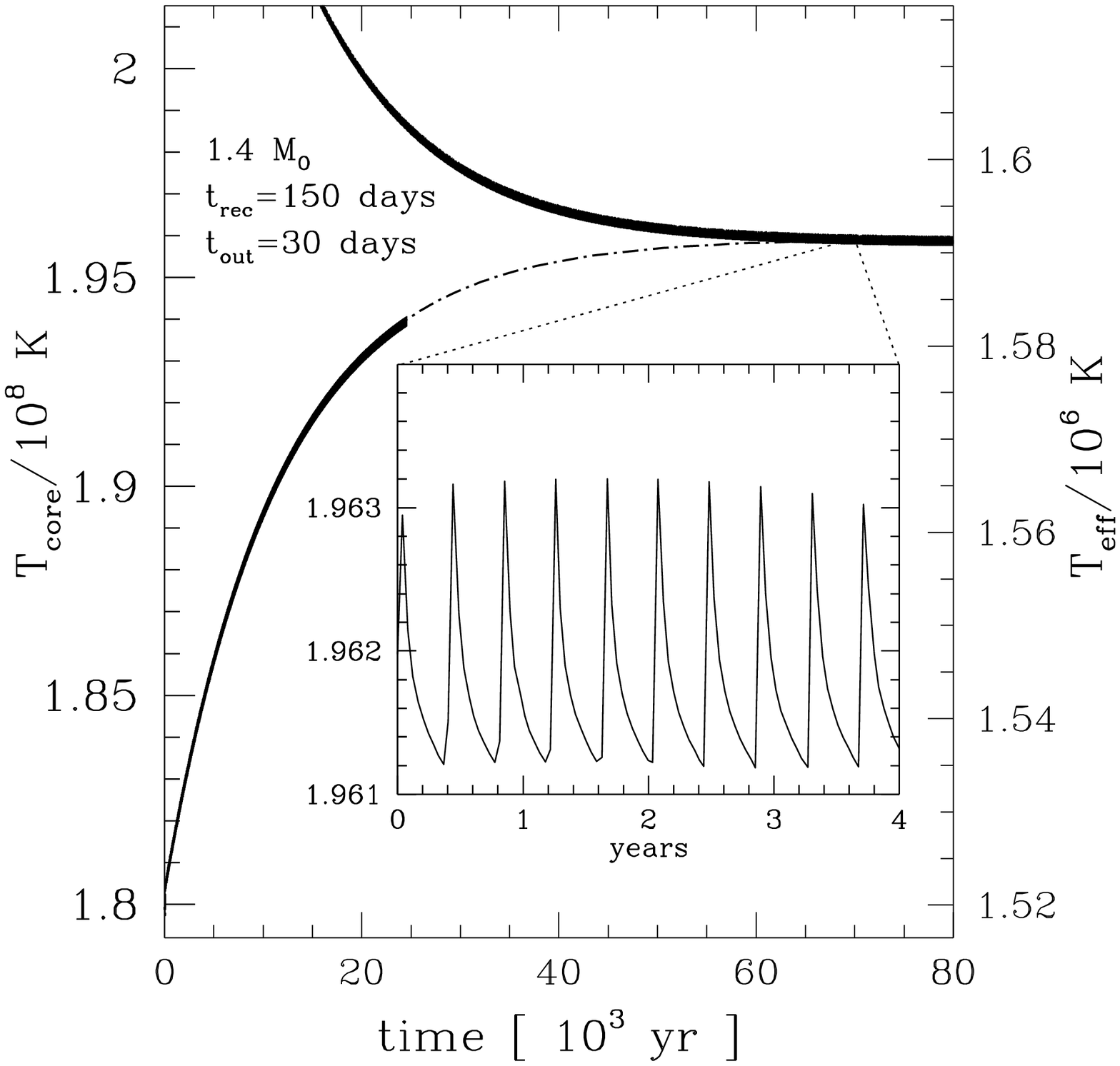}{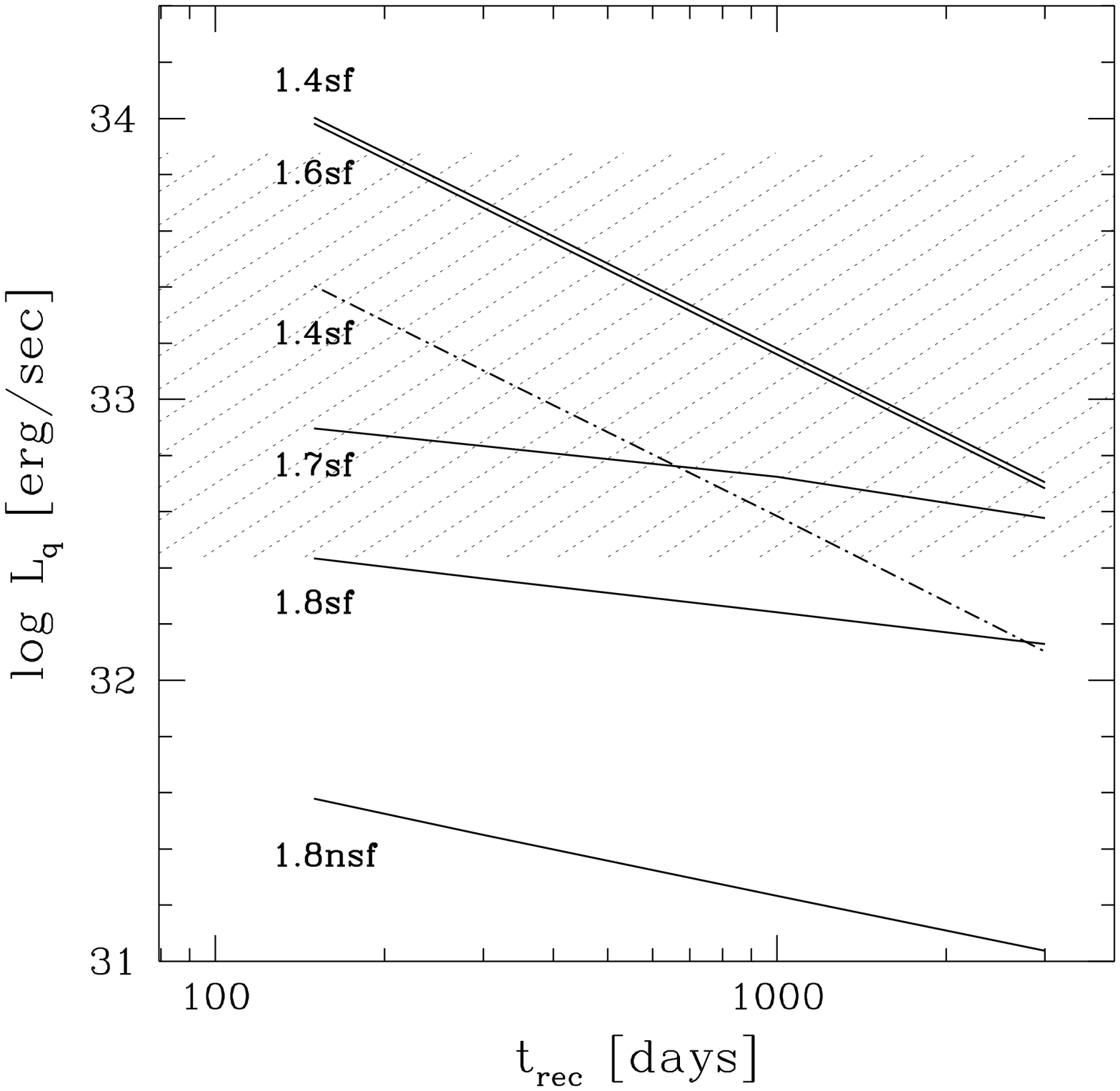}
\caption{
{{\bf a.} Redshifted core temperature $T_{\rm core}$ ($10^8~K$) vs time 
$t$ (yr) soon after the onset of transient accretion, set at $t=0$. The right
scale reports the effective surface temperature as measured at infinity.
The upper curve shows equilibration from a hotter state that
may result from an early phase of steady accretion, whereas the lower curve
mimics a resurrecting accretion episode following a phase of pure cooling. 
The thickness of the lines is because of the rapid
variations due to accretion, as can be seen in the insert
where  $T_{\rm core}$ is plotted against time over a few cycles. 
{\bf b.}
The quiescent luminosity versus the recurrence time
from our calculations. The dashed
area covers the region of the observed luminosity in SXRTs. The solid lines
refer to a star accreting $\Delta M=6\times 10^{-11}$ \Msun per
cycle (compatible with that inferred from the transients Aql X-1, 
4U 1608-522 and Cen X-4). They are labeled by the mass of the star (in solar
masses) and the existence (sf) or non existence (nsf) of a superfluid phase.
The dot-dashed line shows our results for 1.4 \Msun superfluid star 
loading $\Delta M=10^{-11}$ \Msun (it may describe the 
fainter transient SAX J1808.4-3658).}}
\end{figure}

\begin{figure}
\plotfiddle{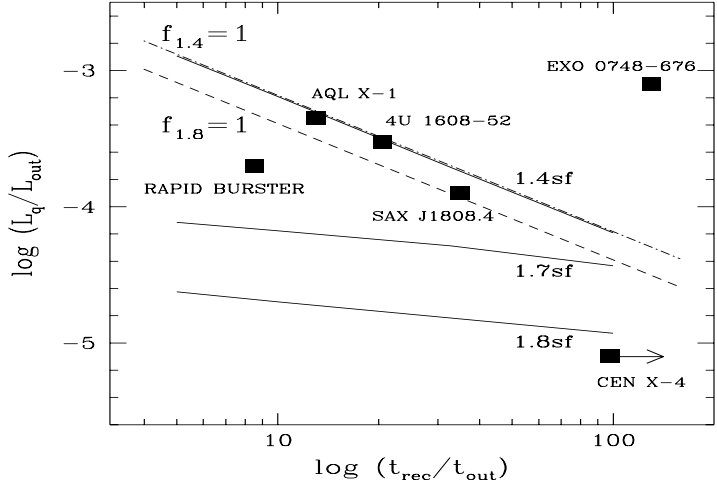}{5.8truecm}{0}{50}{35}{-154}{-55}
\caption{\label{fig:comparison}
{Quiescent to outburst luminosity ratio plotted versus the ratio
of the recurrence time over the outburst time. Bold solid lines are 
the results of our numerical calculations as in Fig.~1b. 
The filled squares represent the observed values.
In the case of Cen X-4 there is still some uncertainty about the value
of $\trec$. The peculiar behavior of EXO~0748-767 can be due to an 
extra luminosity resulting from the energy released by a faint accretion 
and/or by the interaction of the infalling matter with the magnetosphere 
(BBR98).
}}
\end{figure}

We show in Fig.~1a an example of the heating (cooling) of an
initially cold (hot) neutron star after the onset of intermittent accretion. 
It is compared with a model ({\it dot-dashed line}) having constant accretion 
at a rate equal to
$\mdotav=1.46\times~10^{-10}\,\,$\Msun$\!\!{\rm yr}^{-1}.$
The star reaches thermal equilibrium on a time scale 
$\tau_{\rm equ}\sim 10^{4}$ yr, much shorter than any binary evolution time, 
and also much shorter than the time necessary to replenish 
the crust with fresh non catalyzed matter. The equilibration temperature is 
attained when the net injected heat is exactly balanced by the energy loss from
the surface and/or from neutrino emission. 
Fig.~1b shows that $\Lq$ depends crucially on whether fast neutrino emission
in the inner core is allowed or inhibited (and in turn on the value
of the critical temperature $T_c$ for both proton and neutron pairing).
For $1.4-1.6$ \Msun stars with superfluidity (upper solid lines `1.4sf' 
and `1.6sf'), neutrino emission is totally suppressed and the equilibrium 
temperature is determined by balance of nuclear heating with photon cooling. 
When superfluidity is not included, the stars are slightly less luminous
(we omit the results in the figure). In these cases, with a fixed accreted 
mass $\Delta M$ per cycle we see that, naturally, $\Lq$ is 
proportional to $1/\trec$. When fast neutrino emission is allowed 
(`1.7sf'\& `1.8sf' lines), the equilibration temperature is much lower 
and the lowest $\Lq$ are obtained in the case neutrino emission 
is not affected by neutron pairing (`1.8nsf'). In these fast 
cooling models the dependence of $\Lq$ on $\trec$ is weak since 
most of the heat deposited during a cycle is rapidly lost into neutrinos.

\section{Comparison with the data as a tool for probing NS properties}

In Fig.~2 we compare the results of our calculation with the ratios $\Lq/\Lout$
and $\trec/\tout$ observed in 6 SXRTs (Rutledge et al. 2000; Stella et al. 
2000). It appears remarkable that Aql~X-1 and 4U~1608-52 stands just on the 
theoretical line constructed for 1.4 \Msun$\!\!$-star, thus strongly 
supporting the hypothesis that the quiescent luminosity $\Lq$ seen in SXRTs 
comes from the rediffusion toward the surface of the heat deposited in the 
core by the pycnonuclear reactions triggered in the crust due to transient 
accretion (BBR98). $f$ is the fraction of the heat released in the crust 
during accretion which is stored in the stellar interior and later slowly 
leaking out to the surface: the expected results for $f =1$, are shown 
for our 1.4 \Msun $\!\!\!$-star ({\it dashed-dot line}) and 
1.8 \Msun$\!\!\!$-star ({\it dashed line}).
It appears that $f \ll 1$ when direct Urca emission switches on, as in the 
core of our heavier stars (`1.7sf' and `1.8sf', {\it solid lines}).
In these framework we can now use our results to infer properties of the 
neutron star. For example, the measured ratio $\Lq/\Lout$ of Cen X-4 may 
be explained if its NS is heavier than the canonical 1.4 \Msun. 
The Rapid Burster would fit the theory if its mass lies in an intermediate 
range. Measuring the masses of SXRTs by charting the X-ray temperature may 
provide a clue for establishing the link between the NSs in low mass binaries 
and the millisecond pulsars (Stella et al. 1994).
More concretely, combining the observed ratio $\Lq/\Lout$ with future 
independent measurements of the mass of the NSs in the SXRTs, 
one could pose constraints on the value of the critical temperature
for neutron pairing $\Tc$ (Page et al. 2000).


\end{document}